\documentstyle[12pt,sprocl]{article}

\def\beq{\begin{eqnarray}}
\def\eeq{\end{eqnarray}}
\newcommand{\EQ}{\begin{equation}}
\newcommand{\EN}{\end{equation}}
\newcommand{\bea}{\begin{eqnarray}}
\newcommand{\ena}{\end{eqnarray}}
\newcommand{\beas}{\begin{eqnarray*}}
\newcommand{\enas}{\end{eqnarray*}}
\renewcommand{\a}{\alpha}
\renewcommand{\b}{\beta}
\renewcommand{\l}{\lambda}

\newcommand{\G}{\Gamma}

\newcommand{\shalf}{\frac{1}{2}}

\newcommand{\pa}{\partial}
\newcommand{\z}{\zeta}
\newcommand{\ep}{\epsilon}

\begin{document}

\title{KABAT'S SURFACE  TERMS IN THE ZETA-FUNCTION APPROACH }

\author{ D. Iellici \footnote{e-mail address: iellici@science.unitn.it}
 and V. Moretti \footnote{e-mail address: moretti@science.unitn.it} }

\address{Dipartimento di Fisica, Universit\`a di Trento, \\
and INFN Gruppo Collegato di Trento\\
I-38050 Povo (TN), Italy}


\maketitle\abstracts{
The thermal partition functions of photons in any covariant gauge and
gravitons in the harmonic gauge, propagating in a Rindler wedge, are
computed using a local zeta-function approach.
The results are discussed in
relation to the quantum corrections to the black hole entropy.
 The
correct  leading order temperature dependence $T^4$ is
obtained in both cases. For the photons, it  is confirmed the
existence of a surface term giving a negative contribution to the
entropy, as earlier obtained by D.Kabat, but this term is  shown to be
gauge-dependent in the four dimensional case and therefore discarded.
It is argued that similar terms could appear dealing with any integer
spin $s\geq 1$ in the massless case and in more general manifolds. Our
conjecture is checked in the case of a graviton in the harmonic gauge,
where  different surface terms also appear.}

\section{Zeta function approach for photons in the Rindler wedge}

The aim of this work is the computation of the one-loop quantum corrections
to the black-hole entropy due to photons and gravitons. In our approach, the
corrections are identified with the entropy of the quantum fields living
outside the horizon.

In the last years, many papers have dealt with the computation of the one-loop
quantum correction to the entropy of a large mass black-hole, using the
approximation of the Schwarzschild metric given by the simpler Rindler metric
and the conical singularity method.
Most of these works have considered the scalar field only, while,
 last year D.Kabat published
a paper \cite{DK}, 
in which he explicitly considered the fermion and
photon fields.
In the photon case, he found an
unexpected `surface' term in the effective action, in the sense that
it can be interpreted as due to particle paths beginning and ending at
the black-hole surface (horizon). This term makes negative
the corrections to the black-hole entropy at the the Hawking temperature.\\
Kabat employed the heat-kernel plus proper time regularization procedure
which notoriously gives a wrong temperature dependence of the thermodynamical
quantities in dimensions other than two.

For this reason, we investigated the photon case
with the local zeta-function approach developed by Zerbini, Cognola and Vanzo
\cite{zcv}
for the scalar case and which gives the correct temperature behaviour.\\
We have computed the thermal
partition function of a quantum field  employing an
Euclidean path integral over all the field configurations that are
periodic in the imaginary time and identify the period $\beta$ with the
inverse temperature: in doing this, the Rindler wedge becomes
the manifold $C_\beta\times \mbox{R}^2$,where $C_\beta$ is the
cone with angular deficit $2\pi-\beta$.
The one-loop thermal partition function is then given by
$$\ln Z_\b=-\shalf \ln \det L_4+\ln Z_\b^{\mbox{\scriptsize Ghost}}
$$
where, in our case, $L_4$ is the small fluctuation operator of the e.m.
field, that we have considered in any covariant gauge:
$$
L_4=[g^{ab}(-\Delta)+ (1-\frac{1}{\alpha})\nabla^{a}\nabla^{b}],
$$
where $\alpha$ is the gauge fixing parameter and $g^{ab}$ is the Euclidean
Rindler metric:
$$
ds^2=r^2 d\tau^2+dr^2+dy^2+dz^2, \hspace{.5cm}
0<\tau<\b,\,\,\\,\,0<r<\infty,\,\,\,\, y,z\in\mbox{R}
$$
To compute the above determinant, the first step has been to find a
complete set of delta normalized 
eigenfunctions for the small fluctuation operator
on the manifold with conical singularity $C_\b\times \mbox{R}^2$:
\beq
A_a^{(I,n\l{\bf k})}&=&\frac{1}{k}\ep_{ij}\pa^j\phi=
\frac{1}{k}(0,0, ik_z\phi,-ik_y\phi)\nonumber,\\
A_a^{(II,n\l{\bf k})}&=&\frac{1}{\l}\sqrt{g}\ep_{\mu\nu}\nabla^\nu\phi=
\frac{1}{\l}(r\pa_r\phi,-\frac{1}{r}\pa_\tau\phi,0,0),\nonumber\\
A_a^{(III,n\l{\bf k})}&=&\frac{1}{\sqrt{\l^2+{\bf k}^2}}
(\frac{k}{\l}\nabla_\mu-\frac{\l}{k}\pa_i)\phi=
\frac{1}{\sqrt{\l^2+{\bf k}^2}}
(\frac{k}{\l}\pa_\tau\phi,\frac{k}{\l}\pa_r\phi,
-\frac{\l}{k}\pa_y\phi,-\frac{\l}{k}\pa_z\phi),\nonumber\\
A_a^{(IV,n\l{\bf k})}&=&\frac{1}{\sqrt{\l^2+{\bf k}^2}}
\nabla_a\phi=\frac{1}{\sqrt{\l^2+{\bf k}^2}}
(\pa_\tau\phi,\pa_r\phi,\pa_y\phi,\pa_z\phi), \nonumber
\eeq
where $\phi=\phi_{n\l{\bf k}}(x)$ is the complete
set of delta normalized eigenfunctions of the  Friedrichs self-adjoint
extension of the scalar Laplacian on $C_\b\times \mbox{R}^2$ with eigenvalues
$-(\l^2+{\bf k}^2)$:
\beq
\phi_{n\l{\bf k}}(x)&=&\frac{\sqrt{\l}}{2\pi\sqrt{\b}}e^{ik_y y+ik_z z}
e^{i\frac{2\pi n}{\b}\tau} J_{\nu_n}(\l r),\hspace{5mm}
n=0,\pm 1, \dots;\,\,\l\in \mbox{R}^+;\,\, k_y, k_z\in \mbox{R}
\nonumber
\eeq
The first three eigenfunctions satisfy $\nabla^a A_a=0$
and have eigenvalue $\l^2+{\bf k}^2$, while $A_a^{(IV)}$ is a pure
gauge and has eigenvalue $\frac{1}{\a}(\l^2+{\bf k}^2)$.\\
Using these modes we have then written the local zeta-function of the e.m.
field using its spectral representation:
\beq
\z(s;x)=\sum_i\sum_n\int d\l\,d^2{\bf k}\; [\nu_i^2(n\l{\bf
k})]^{-s} g^{ab} A_a^{(i)}(x)A_b^{(i)\ast}(x)\nonumber\:.
\eeq
Since the manifold is non-compact, only the local zeta-function is directly
definable
as a finite quantity;
 to define the global zeta-function one needs to introduce some
smearing function in the spatial integrations.\\
After some manipulations, we have found that the zeta-function can be
expressed in terms on the zeta-function of a minimally coupled scalar
field and a new term which is a total derivative:
$$
\z(s;x)=(3+\a^s)\z_\b^{\mbox{\scriptsize Scalar}}(s;x)+
\frac{s+1+\a^s(s-1)}{2(s-1)}\Delta
\z_\b^{\mbox{\scriptsize Scalar}}(s+1;x).
$$
where the scalar local zeta-function in this background has been
computed by Zerbini, Cognola and Vanzo \cite{zcv}:
$$
\z_\b^{\mbox{\scriptsize Scalar}}(s;x)=\frac{r^{2s-4}}{4\pi\b\G(s)}
I_\b(s-1),
$$
where $I_\b(s)$ is a function analytic in the whole complex plane but in
$s=1$, where it has a simple pole, and its value in the interesting
points $s=0,-1$ are known \cite{zcv}.\\
After adding the contribution of the ghosts \footnote{This contribution follows
by a direct zeta-function approach on the gauge fixing 
Faddeev-Popov determinant
(which appears in the complete photon Feynman integral)
 taking its $\alpha-$dependence
into account.},
which is  $-2\a^{(s/2)}\z_\b^{\mbox{\scriptsize Scalar}}(s;x)$,
and using the formulae
\beas
{\cal L}_\b(x)&=&\shalf \z'(s=0;x)+\shalf\z(s=0;x)\ln
\mu^2,\nonumber\\
\ln Z_{\b}&=&\int d^4 x \sqrt{g}\,{\cal L}_\b(x),
\enas
we find that the one loop effective Lagrangian density for the
electromagnetic field on $C_\b\times \mbox{R}^2$ is
\beas
{\cal L}_\b^{\mbox{\scriptsize e.m.}}(x)&=&2{\cal
L}_\b^{\mbox{\scriptsize
 Scalar}}(x) -\frac{(1-\shalf\ln\a)}{2\pi\b r^4}I_\b(0)\nonumber\\
&=&\frac{1}{4\pi\b r^4}I_\b(-1)-\frac{(1-\shalf\ln\a)}{2\pi\b
r^4}I_\b(0).
\enas
Some comments on the found results are in order.\\
The twice-scalar part is expected from the physical ground.
 It is also clear that ${\cal L}_\b^{\mbox{\scriptsize e.m.}}(x)$ may depend
on the gauge fixing parameter $\a$ since the
gauge invariance must hold for integrated quantities like the
effective action \footnote{One can notice that the two-dimension case is
 misleading, since
the one-loop effective lagrangian density is indeed gauge-invariant.}.\\
Anyway, when integrated, the gauge-dependent term gives rise to
Kabat-like surface term and gives a negative contribution to the
entropy, at least for some values of $\a$.
However, since the gauge-dependent part take rise form a total
derivative, such a term would be discarded after the integration over
the space-time if this were a smooth, compact manifold without borders
or singularities: this is a standard procedure when checking the gauge
invariance of a theory.
 As a consequence, our procedure to restore the gauge invariance
is simply to discard the `surface' term: in this way we also avoid
embarrassing negative entropies. This procedure reflects the fact that
local quantities like the the local zeta-function are ill-defined due
to the possibility of adding a total derivative with vanishing
integral.\\
After discarding the `surface' term, we can compute the thermodynamical
quantities in the usual way: for example  the renormalized free energy,
the energy density and the entropy are:
\beas
F_\b^{\mbox{\scriptsize Sub.}}&=&-\frac{A_\perp\pi^2}{90\b^2\ep^2}
-\frac{A_\perp}{36\b^2\ep^2}-\frac{13 A_\perp}{1440\pi^2\ep^2},\\
<T_0^0>^{\mbox{\scriptsize Sub.}}&=&\frac{\pi^2}{15\b^4 r^4}+
\frac{1}{18\b^2 r^4}-\frac{13}{720\pi^2 r^4},\\
S_\b&=&\b^2\pa_\b F_\b=\frac{A_\perp}{90\b\ep^2}
\left [\left(\frac{2\pi}{\b}\right)^2+5\right].
\enas
These results are in agreement with twice the minimally coupled
scalar field results obtained with different methods (point-splitting,
optical metric) \cite{DC,BO,DO94}
but for the coefficient of the term $\b^{-2}$ for which
we get one third of their result. This discrepancy appears also in the
heat-kernel approach \cite{CKV} and is not
yet understood.

\section{A general conjecture  and the graviton case}

Investigating the origin of the surface gauge-dependent term
 we have conjectured the appearence of such terms on more general
manifolds in the form ${\cal {M}}\times\mbox{R}^2$, 
where the generally curved
manifold ${\cal {M}}$
has conical singularities or boundaries, and for higher spin fields.
The general form of such terms as far as the total photon 
zeta function contribution 
is concerned  should be there:
\begin{eqnarray}
\zeta^{\scriptsize \mbox{surface}}_{{\cal M}\times 
\mbox{\scriptsize R}^2}(s;x)
= \frac{s+1+\alpha^s(s-1)}{2s}\:
   \frac{\Gamma(s-1)}{4\pi \Gamma(s)} \nabla_a \sum_n \int d\lambda
\lambda \,{\bf J}^\ast \nabla^a {\bf J}\:.\nonumber
\end{eqnarray}
where ${\bf J}_{n,\lambda}(x^\mu)$ is an eigenfunction of (the Friedrichs
extension of) the 0-forms Hodge Laplacian on ${\cal M}$.\\
We have checked our hypothesis in the case of the graviton field
in the Rindler wedge employing the harmonic gauge,
 and we have indeed found many and more
complicated surface-like terms, dropping which one gets
the expected physical result for the effective action:
$$
\ln Z_\b^{\mbox{\scriptsize Gravitons}}=
2 \ln Z_\b^{\mbox{\scriptsize Scalar}}(s;x).
$$

\section*{Acknowledgments}

We would like to thank L. Vanzo and S. Zerbini for valuable discussions 
on several topics studied in this paper and G. Esposito for very useful
suggestions.

\end{document}